\documentclass[aps,pre,groupedaddress,amssymb,amsmath,pra,floatfix,showpacs]{revtex4}
\usepackage{epsfig}

\begin{document}
\title{Identification of GW bursts in high noise using Pad\'{e} filtering}
\author{L. Perotti${}^{*}$, T. Regimbau ${}^{\dag}$, D. Vrinceanu${}^{*}$ and D. Bessis${}^{*}$}
\affiliation{${}^{*}$Department of Physics, Texas Southern University, Houston, Texas 77004 USA}
\affiliation{${}^{\dag}$ UMR ARTEMIS, CNRS, University of Nice Sophia-Antipolis,
Observatoire de la C\^{o}te d'Azur, BP 4229, 06304, Nice Cedex 4, France}
\date{\today}

\begin{abstract}

We consider the case of highly noisy data coming from two different antennas, each data set containing a damped signal with the same frequency and decay factor but different amplitude, phase, starting point and noise. 
Formally, we treat the first data set as real numbers and the second one as purely imaginary and we add them together. This complex set of data is analyzed using Pad\'{e} Approximations applied to its Z-transform. Complex conjugate poles are representative of the signal; other poles represent the noise and this property allows to identify the signal even in strong noise. The product of the residues of the complex conjugate poles is related to the relative phase of the signal in the two channels and is purely imaginary when the signal amplitudes are equal. Examples are presented on the detection of a fabricated gravitational wave burst received by two antennas in the presence of either white or highly colored noise.

\end{abstract}


\maketitle

\section{Introduction}

In many fields of physics, one of the primary challenge is to extract a signal buried into noise. This is particularly true for gravitational-wave astronomy, for which the signal is expected to be very weak compared to the instrumental noise of the GW detectors.

Ground based GW interferometers (LIGO \cite{LIGO}, Virgo \cite{Virgo}, GEO600 \cite{GEO600} and TAMA300 \cite{TAMA300}), in operation during the first decade of the century, have demonstrated the feasibility of the experiment. The second generation of LIGO/Virgo detectors (Advanced LIGO \cite{Ad LIGO}, Advanced Virgo \cite{Ad Virgo}) will start collecting data in 2015, opening the era of the first detections, including tens of compact binary coalescences a year and maybe occasional supernovas.

The first sources to be detected are expected to be very weak and much effort has been devoted to the development of powerful data analysis method in order to optimize the chance of an early detection.  

The optimal detection strategy is to apply match filtering but it assumes that one should know the signal, at least with some level of accuracy. This technique is used for the search for compact binary coalescences \cite{burst}, but cannot be applied to the search for un-modeled bursts, produced for instance by core-collapse supernova or the merger of two neutron stars or black holes. Also it is usually computationally very costly as one as to scan the large multi-dimensional parameter space, and one may resort to sub-optimal methods, usually based on the search for excess power in the time/frequency domain \cite{timfre1,timfre2}.

In this paper, we propose an original method based on Pad\'{e} Approximation applied to Z-transform. As an example of its application, we use a simple but realistic model of a ring down which is present in coincidence in two separated detectors with different response functions.

Our method stems from the observation that while the poles of the Pad\'{e} approximant to the Z-transform of a time series of real numbers always happen in complex conjugate (c.c.) pairs, this is not in general true for a series of complex data, with the noticeable exception of poles corresponding to signals that appear both in the real and imaginary part of the data sequence and are not too weak, i.e. their residues are at least comparable to those of nearby poles. Since the residues of noise poles go in average as $1/\sqrt{N}$, where $N$ is the number of data points used to build the Pad\'{e} approximant, while the residues of signal poles do not depend on $N$, this latter condition is not overly stringent.

The idea is therefore to combine the time series output of two antennas into a single time series of complex numbers, calculate the poles of the Pad\'{e} approximant to its Z-transform, and use the above property as a signature of the presence of a signal in coincidence in the two antennas.

This gives us a simple test of the presence of the same signal in the two antennas by detecting the presence of the same (dominating) frequency in both detectors.

In the case of just two channels, the method has the added advantage of being fast as compared to running separate analysis of the two channels and then comparing the results. For higher number of channels the signals from the different channels can be coupled two by two or a matrix Pad\'{e} formalism can be used \cite{beck} and this advantage is lost in part.
 
We note here that as our method concentrates on the position of the poles of the Z-transform of the data series, it depends on the main difference between the the Pad\'{e} approximant and the Fourier transform: the former tries to locate the poles of the Z-transform in the complex plane, while the latter looks at the value of the Z-transform itself at fixed points on the unit circle. This difference is also the source of the so called super-resolution property of the Pad\'{e} approximant \cite{superr,super2}.
 
\section{The theoretical background and the model}

We consider the case of two  GW antennas whose time series output is the sum of the instrumental noise $n(t)$ and a common GW signal $h(t)$ :
\begin{equation}
s_i(t)=n_i(t) + h_i(t) \,\ i=1,2
\end{equation} 
The GW signal at the output of the two antennas is given by :
\begin{equation}
h_i(t)=F_{+,i}(t,ra,\delta,\psi) h_+(t) + F_{\times,i}(t,ra,\delta,\psi) h_{\times}(t)  
\end{equation} 
where $h_+$ and $h_{\times}$ are the plus and cross polarizations in the reference frame of the source, and $F_+$ and $F_{\times}$ the angular pattern functions of the detector that depend on the relative orientation of the source and the detector. In this expression, $ra$ is the right ascension, $\delta$ the declination and $\psi$ the polarization.

Considering the case of a damped sinusoidal we can write :
\begin{equation}
h_i(t)=2A_i \mathrm{exp}^{-\alpha t} \cos(\omega t+\varphi_i) \qquad A_i\quad real, \quad i=1,2.
\end{equation}
In the case of non coincident and non aligned detectors, both the amplitudes $A_i$ and the phases $\varphi_i$ are different, while the frequency and the damping time are the same in the two detectors.
We assume there is no correlation between the noise in the two detectors.

It is the most basic model of a ring-down signal, but since most models have a strongly dominant frequency and their initial anti-causal part is very short (see for example Ref. \cite{imbh} and the expected core-collapse supernova signals in Ref. \cite{super}), it is sufficient for now, especially  since our aim is not description of the signal but detection of its presence in the given data sequence.

In our algorithm, out of the two real signals above, we build a single complex one:
\begin{equation}
s(t)=s_1(t)+is_2(t)
\end{equation}

We assume both signals to be probed at the same regular time intervals $jT$, thus obtaining the data series $s_{0},s_{1},s_{2},.....s_{j},.......$, where $s_{j}=s(jT)$. If we now apply the Z-transform algorithm \cite{couch}
\begin{equation}
Z(z)=\sum_{j\geq 0}s_{j}z^{j},  \label{1}
\end{equation}
to the total complex signal, we get {\it in the case no noise is present}
\begin{equation}
Z_{s}(z)=\frac{A_1e^{i\varphi_1}+iA_2e^{i\varphi_2}}{1-ze^{(-\alpha +i\omega )T}}+\frac{A_1e^{-i\varphi_1}+iA_2e^{-i\varphi_2}}{1-ze^{^{(-\alpha -i\omega )T}}}  \label{1b}
\end{equation}

The two poles of the Z-transform appear in a c.c. pair and the product of their residues is:
\begin{equation}
\rho _{+}\rho _{-}=(A_1e^{i\varphi_1}+iA_2e^{i\varphi_2})(A_1e^{-i\varphi_1}+iA_2e^{-i\varphi_2})e^{-2\alpha T}=[(A_1^{2}-A_2^{2})+i2A_1A_2\cos (\varphi_1 -\varphi_2)]e^{-2\alpha T}  \label{2}
\end{equation}
which is in general any complex number, with the following exceptions:

{\bf 1)} either $A_1$ or $A_2$ is zero: in this case the product is real;

{\bf 2)}  $A_1$ and $A_2$ are equal: in this case the product is imaginary;

{\bf 3)} the difference of phase $\varphi_1 -\varphi_2$ is an odd multiple of $\frac{\pi }{2}$, i.e the signals are in quadrature: in this case the product is again real; this last case is of probability zero.

Extension to any finite number of damped sinusoidal signals in the two channels is a trivial matter of summation.

For a data series which is truncated at $j=N-1$, the sum in eq. \ref{1} is equally truncated; we therefore construct its sub-diagonal Pad\'{e} Approximant, i.e. a rational function with the numerator having the same degree as that of the denominator minus one and whose Taylor expansion equals the Z-transform up to order $N-1$ \cite{pade,dou4,baker,noi}. The aim of this construction is to try and predict the Z-transform for $N \to \infty$. The choice of the sub-diagonal Pad\'{e} Approximant is dictated by the form of the Z-transform for a finite number of poles as can be seen from eq. \ref{1b}.

If a finite number $m$ of frequencies is present in both antennas, $N$ equals four times their number, and there is no noise, the poles of the Pad\'{e} Approximant to the truncated Z-transform again appear in c.c. pairs; if all the above conditions do not hold true the poles do not in general appear in c.c. pairs.

To see this, let us consider the simplest case: a signal of the form
\begin{equation}
2(1+i)\cos (\omega t) + C, \qquad C\in \mathbb{R},
\end{equation}
which we sample at every quarter of period, so that its Z-transform is   
\begin{equation}
Z_{s}(z)=\frac{1+i}{1-iz}+\frac{1+i}{1+iz}+\frac{C}{1-z}.
\end{equation}
We now truncate the time series after $N=4$ points and construct the [1/2] Pad\'{e} Approximant; its denominator is 
\begin{equation}
Q(z)=(s_2^2-s_1s_3)z^2+(s_0s_3-s_1s_2)z+(s_1^2-s_0s_2).
\end{equation}
where $s_0=2(1+i)+C$, $s_2=-2(1+i)+C$ and $s_1=s_3=0$, so that
\begin{equation}
Q(z)=s_2(s_2z^2-s_0) 
\end{equation}
whose roots 
\begin{equation}
z_{\pm}=\pm \sqrt{\frac{C+2(1+i)}{C-2(1+i)}} \label{4}
\end{equation}
are not in general complex conjugate.

Back to the general case of $m_1$ and $m_2$ frequencies in the two channels, we have the following cases:

{\bf 1)} If the signal is only in the first (or second) antenna, then we have c.c. poles even if $N$ is less than four times the number of frequencies $m_1$ (or $m_2$).
 
{\bf 2)} If the two antennas carry the same frequencies, we again have c.c. poles even if $N$ is less than four times the number of frequencies $m_1=m_2$, but only if the ratios of amplitudes in the two channels are the same for all frequencies. 

{\bf 3)} If the two antennas carry equal numbers of different frequencies and $N$ is less than four times the total number of frequencies, we do not in general have c.c. poles.

{\bf 4)} If the two antennas carry different numbers of different frequencies and $N$ is less than four times the total number of frequencies, we again have in general non c.c. poles.

On the other hand, poles corresponding to strong signals dominate the spectrum by strongly repelling other poles and still appear in almost c.c. pairs. As an example we see that when $C\ll 1$ the two roots given by eq. \ref{4} above become $z_{\pm}\simeq \pm \sqrt{-1+C(1-i)/2}$ which are very close to $\pm i$.

If noise is not too strong it is therefore possible to detect coincident signals by looking for c.c. pairs of poles; checking that the product given by eq. \ref{2} is the same when calculating the Pad\'{e} Approximant on a sample starting from a different point within the data sequence allows for an extra check of the presence of coherent signals in the two antennas. The numerical advantage is that instead of calculating two Pad\'{e} Approximants, we only have to calculate one.

The presence of two coincident signals in the two antennas can therefore be deducted from eq. \ref{2}. Notice that the phases and amplitudes of the two signals cannot be obtained this way, since the procedure only gives us a single complex number for the combined residue. On the other hand, they can be easily obtained from 
\begin{eqnarray}
\frac{\rho _{+}}{z_+}+\left({\frac{\rho _{-}}{z_-}}\right)^*=2A_1e^{i\varphi_1}\\
\frac{\rho _{+}}{z_+}-\left({\frac{\rho _{-}}{z_-}}\right)^*=2iA_2e^{i\varphi_2}
\end{eqnarray}
where $z_+$ and $z_-$ are the positions of the poles.

\section{Numerical Results}

We give here a few examples of the numerical implementation of our method over a toy model of a ring-down GW burst. The model assumes a sampling rate $sr=1024Hz$, a ring-down with period of $10ms$ and damping time $\tau=0.1s$, corresponding to the formation of a Intermediate-mass black hole (IMBH) with dimensionless spin parameter $\hat{a}=0.999835$ and mass $M=305$ solar masses \cite{imbh}. We note here that what matters in our method are not the times themselves but their value in terms of the sampling ratio, i.e. the number of data points spanned by a period of the signal ($10$ points here) and by the signal itself (a few hundred data points).

\subsection{Weak white Gaussian noise}  

We start with two antennas receiving the same signal and having as output the signal shown in Figure \ref{fig01}; to this we add white Gaussian noise with standard deviation $0.1$ times the maximum amplitude of the signal. Following the standard practice when dealing with transient signals, we analyze the signal over a traveling window. The window length is $100$ data points and its step is $2$ data points; the range of the first and last window is indicated by the two double arrows in the figure. The first window in which the signal appears is window number $20$; the window for which the signal starts at the beginning of the window itself is window number $70$. Figure \ref{fig02} shows as black crosses the angular position $\phi$ in radians of poles in the upper half of the complex plane and as red St. Andrew ones that of the c.c.'s of poles in the lower half; the actual frequencies being $f=sr\times\phi/2\pi$. The signal is clearly visible as a horizontal line of superimposed black and red crosses. Note that the line only starts shortly after window number $45$, that is: it only appears when the signal starts before the mid of the window. This is due to the fact that the z-transform of a delayed signal is not a rational function of the form $[(M-1)/M]$ with $M$ an integer: for a single transient damped signal, starting at some point $n_{0}\geq 1$, 
\begin{eqnarray}
s_{n}=0\qquad\qquad\qquad\qquad\qquad\qquad\qquad\qquad\qquad\qquad\quad 0\leq n<n_{0}\nonumber\\
s_{n}=2A\mathbf{Re}\left[{e^{i\omega\frac{T}{N}(n-n_{0})}}\right]\qquad \omega =2\pi \nu+i\alpha\qquad \alpha>0\qquad n\geq n_{0},
\end{eqnarray}
the $Z-transform$ reads:
\begin{equation}
Z_{S}(z)=z^{n_{0}}\left({\frac{A}{1-ze^{i\omega\frac{T}{N}}}+\frac{A}{1-ze^{-i\omega^{\ast }\frac{T}{N}}}}\right).
\end{equation}
which is a rational function of the type $[(n_0+1)/2]$. More in general, for $K$ (damped) sinusoidal signals, all starting at point $n_0$ the Z-transform is a rational function of the type $[(n_0+2K-1)/2K]$. A Pad\'{e} Approximant of the form $[(M-1)/M]$ is not therefore suited (even with $M=2K$) to reconstruct it and completely fails to detect the signal if it starts after the mid of the time interval on which the approximant is calculated, even for high $M$ and in the presence of noise.

As the signal is large, its poles strongly repel other poles \cite{barone3,barone}, as we can see in figure Figure \ref{fig02b} where -instead of using a single noise realization whose starting point is shifted step by step- the whole noise realization is changed at each window: this way we have on the plane of the figure a uniform distribution of the noise poles from which repulsion from the signal poles clearly stands out. The repulsion slightly decreases with increasing window number, as the peak amplitude of the signal within the window decreases.

\subsection{Weak colored noise}

We now pass to a more realistic noise model, built by multiplying the spectrum of Gaussian white noise with zero mean and unit variance times the square root of the noise power spectral density (PSD) of a given detector \cite{rod} and then inverse Fourier transforming the result to obtain the time series. The PSD's used are the nominal ones for advanced LIGO and Virgo and are truncated at $10Hz$.

Using the Hanford PSD for the first channel and the Livingston one for the second,
 we get the picture in Figure \ref{fig04}: the straight line is still clearly visible and again there is very little correspondence between noise poles in the upper half of the complex plane and those in the lower half, except when the curves happen to cross.

This behavior suggest the following scheme of detection: we search couples of c.c. poles whose elements are distant in the complex plane less than a given value $\delta_1$ and which repeat for at least two (or eventually more) windows within a distance $\delta_2$. For these poles we calculate the quantity
\begin{equation}
\frac{Im(\rho _{+}\rho _{-})}{|\rho _{+}\rho _{-}|}\in [-1,1]\label{3}
\end{equation}
which is zero in cases 1) and 3) above and $\pm 1$ if $A_1=A_2$. A constant value of the quantity eq. \ref{3} indicates that the signals in the two channels remains in phase. The result for the poles of Figure \ref{fig04} is shown in Figure \ref{fig05}.

\subsection{A different couple of antennas}

If instead we consider for the two channels the two transmitted signals of Figure \ref{fig06}, and for the noise PSD's the Hanford and Virgo ones, the result is that of Figure \ref{fig07}: the quantity (\ref{3}) is around $-0.5$. The choice $\delta_1=\delta_2=0.01$ allows us to identify only about the first half of the signal, which on the other hand is clearly visible by eye. This suggests that their value will need to be increased when increasing the noise level, taking care of keeping both $\delta_1$ and $\delta_2$ smaller than the average spacing of the poles $4\pi/N$ which for $N=100$ is $0.125$.

\subsection{Stronger noise}

If the noise level is increased, more and more noise poles arrange themselves in c.c. pairs while the separation between the poles of a signal c.c. pair increases and the picture we gave above becomes less clear. The important parameter appears to be the amplitude of the smaller of the two signals: when it becomes comparable to the noise standard deviation, detection becomes problematic.

\begin{figure}[htbp]
\centering\epsfig{file=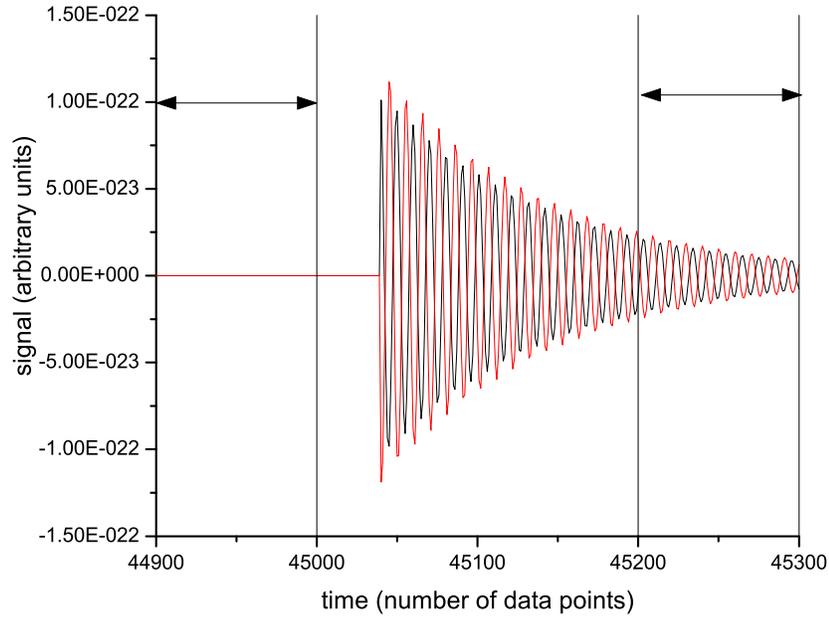,width=0.7\linewidth}
\caption{The output of two antennas receiving the same damped sinusoidal signal (black in the first channel, red in the second one). Amplitudes are equal and phases are opposite. The signal starts after about 140 data points and lasts for about 500 data points.}
\label{fig01}
\end{figure}

\begin{figure}[htbp]
\centering\epsfig{file=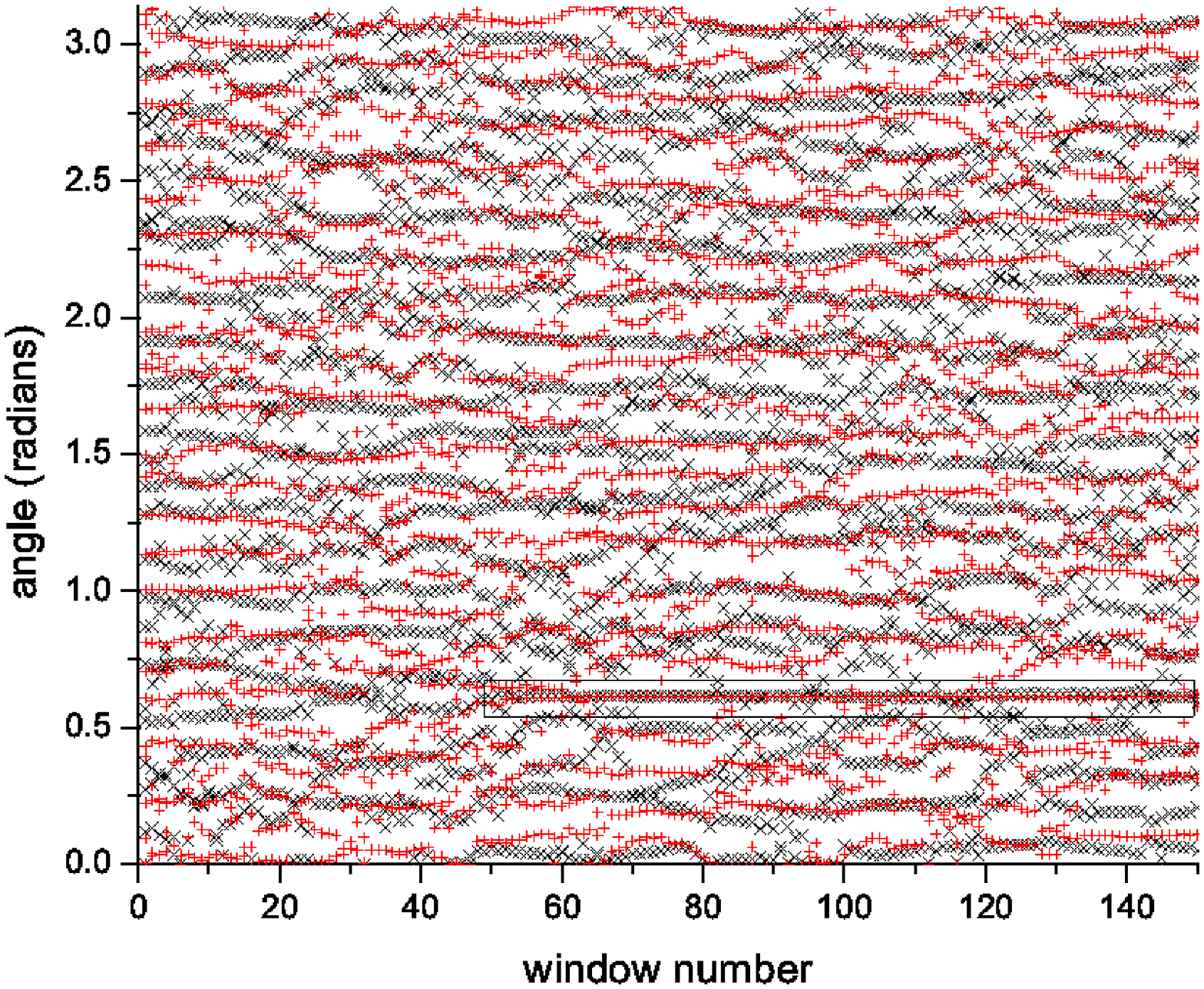,width=0.7\linewidth}
\caption{Poles in the upper half of the complex plane (black crosses) and c.c.'s of the poles in the lower half of the complex plane (red crosses) for the signals in Figure \ref{fig01} in white Gaussian noise whose standard deviation in each channel is $0.1$ times the maximum amplitude of the signal in the first channel. Traveling window of length $100$ data points, with step $2$ data points.}
\label{fig02}
\end{figure}

\begin{figure}[htbp]
\centering\epsfig{file=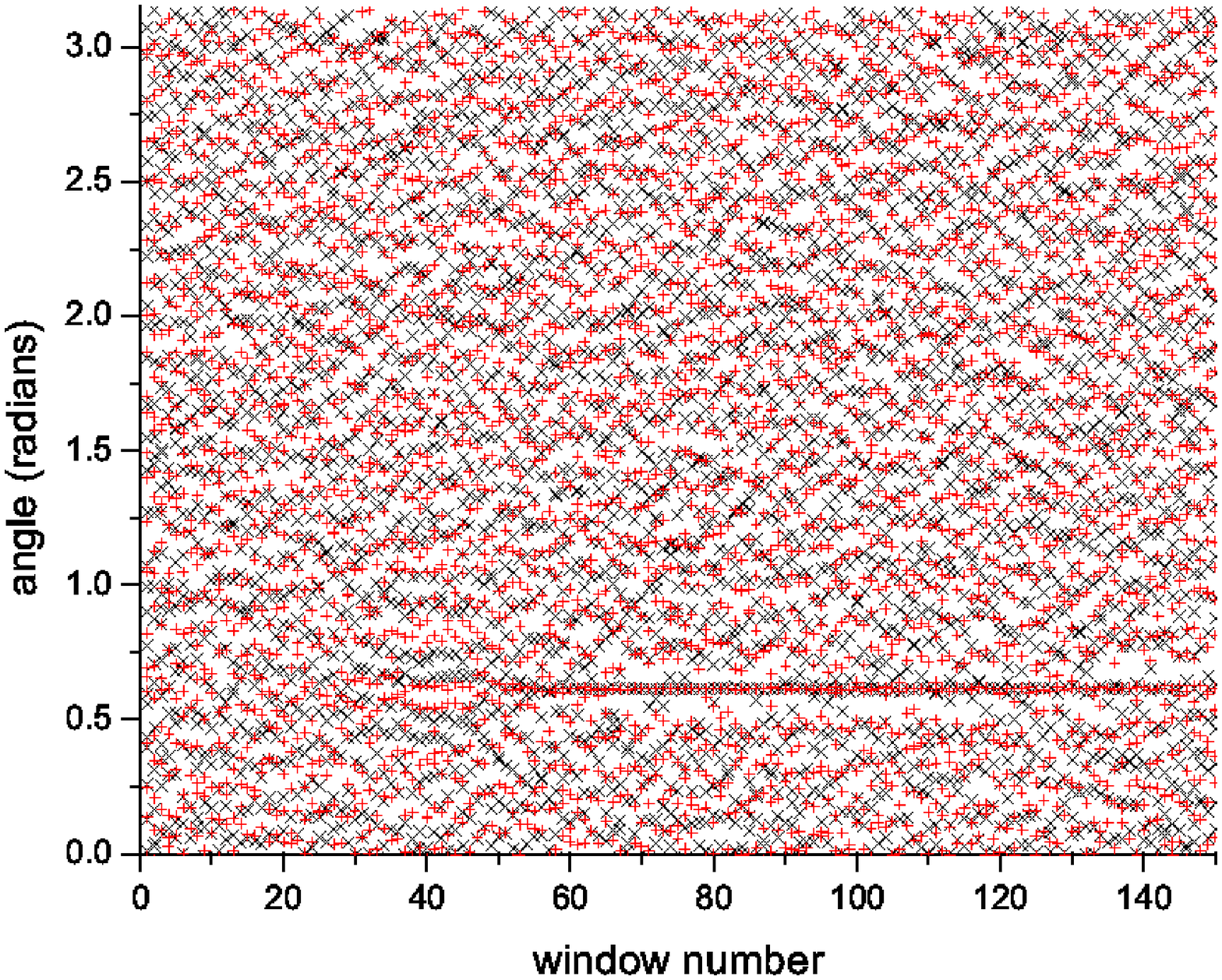,width=0.7\linewidth}
\caption{Poles in the upper half of the complex plane (black crosses) and c.c.'s of the poles in the lower half of the complex plane (red crosses) for the signals in Figure \ref{fig01} in white Gaussian noise whose standard deviation in each channel is $0.1$ times the maximum amplitude of the signal in the first channel. Traveling window of length $100$ data points, with step $2$ data points. The noise realization is changed at each window.}
\label{fig02b}
\end{figure}

\begin{figure}[htbp]
\centering\epsfig{file=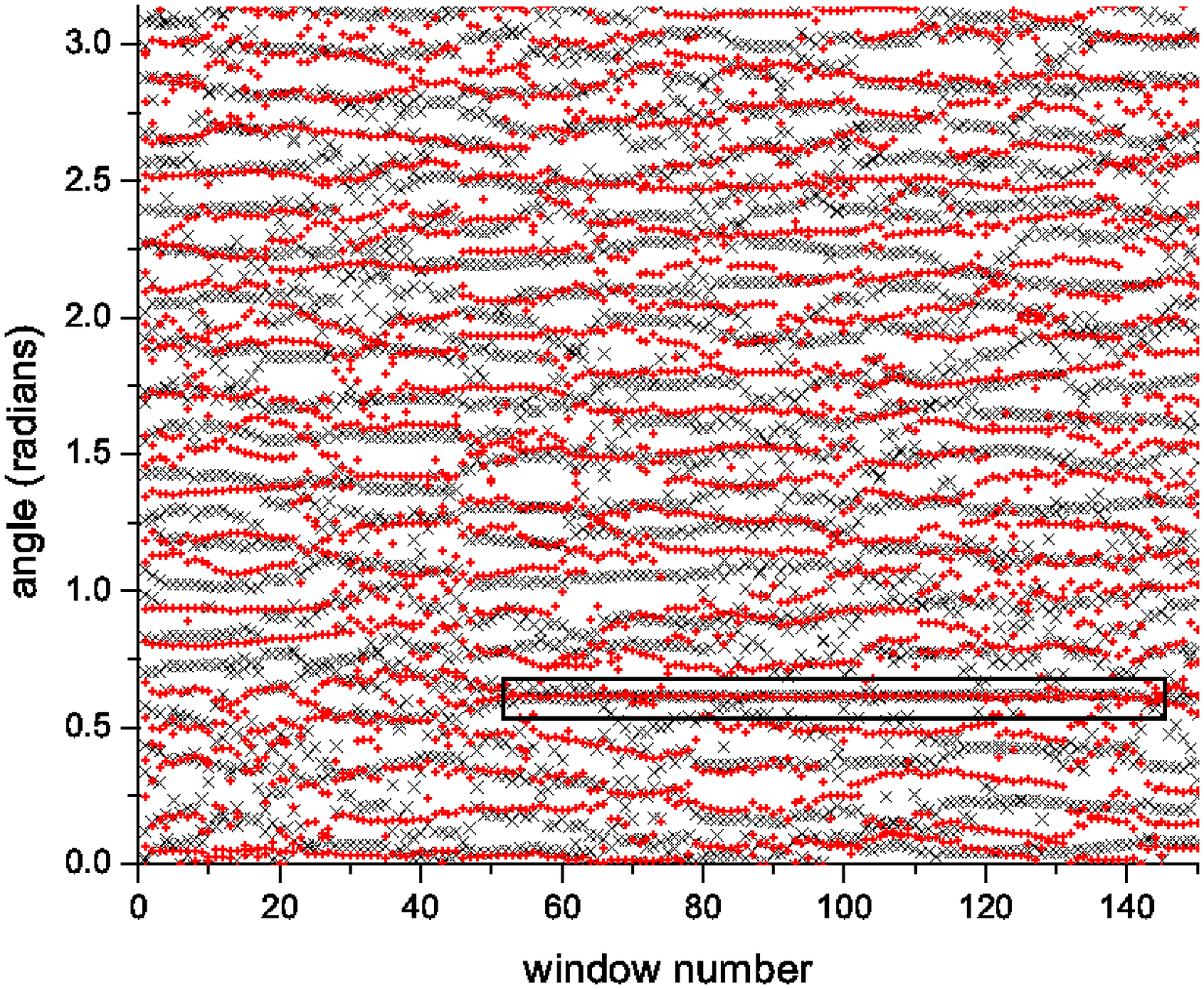,width=0.7\linewidth}
\caption{Poles in the upper half of the complex plane (black crosses) and c.c.'s of the poles in the lower half of the complex plane (red crosses) for the signal in Figure \ref{fig01} in the Hanford and Livingston colored noise. 
Its standard deviation in each channel is $0.36$ times the maximum amplitude of the signal in the first channel. Traveling window of length $100$ data points, with step $2$ data points.}
\label{fig04}
\end{figure}

\begin{figure}[htbp]
\centering\epsfig{file=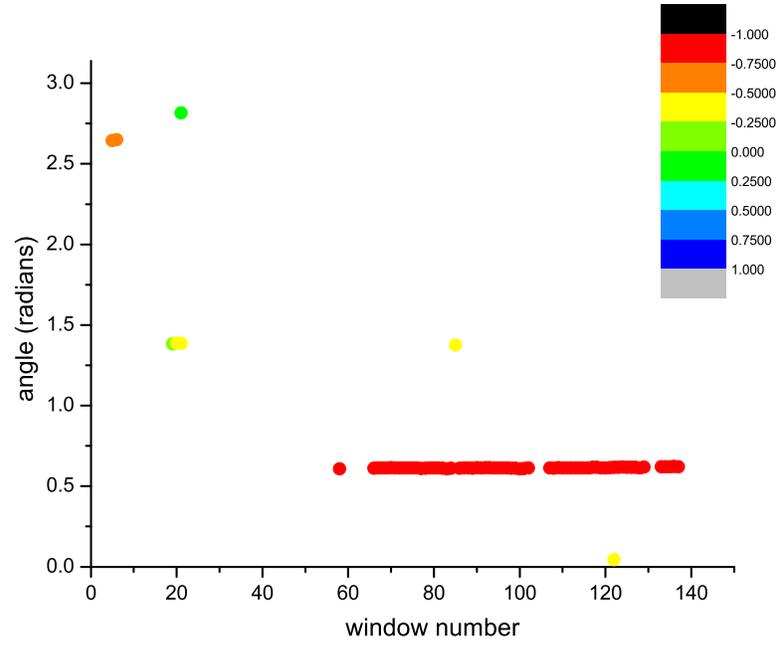,width=0.7\linewidth}
\caption{Coincidences of c.c. poles from Figure \ref{fig04}. $\delta_1=\delta_2=0.01$. The color of the dots represents the quantity eq. \ref{3}.}
\label{fig05}
\end{figure}

\begin{figure}[htbp]
\centering\epsfig{file=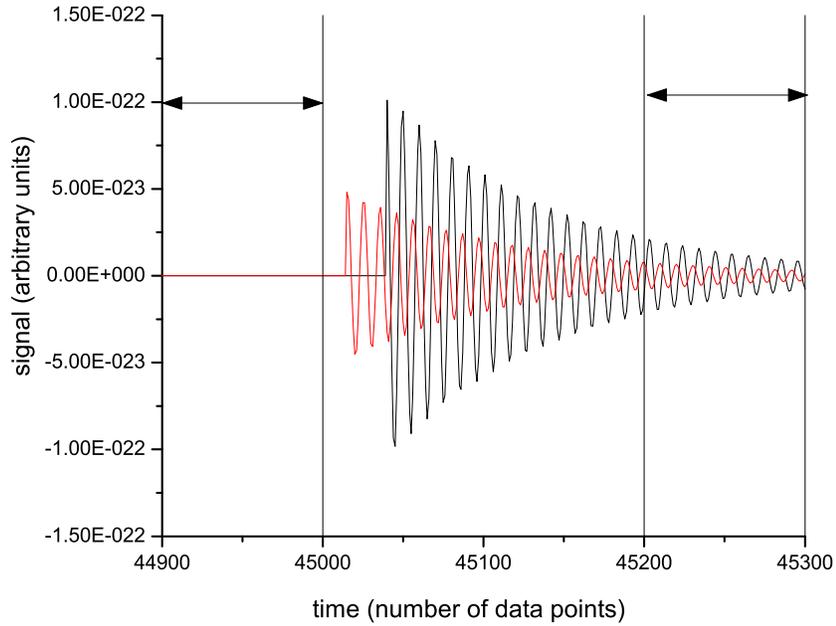,width=0.7\linewidth}
\caption{The output of two antennas receiving the same damped sinusoidal signal (black in the first channel, red in the second one). Both amplitudes and starting points are different; the signals are almost in quadrature.}
\label{fig06}
\end{figure}

\begin{figure}[htbp]
\centering\epsfig{file=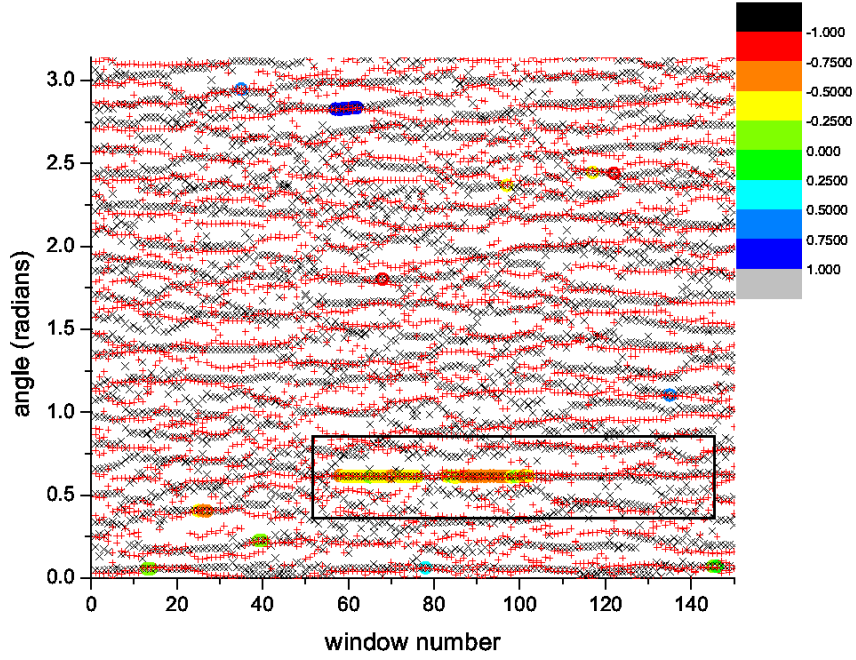,width=0.7\linewidth}
\caption{Poles in the upper half of the complex plane (black crosses), c.c.'s of the poles in the lower half of the complex plane (red crosses), and coincidences of c.c. poles (colored circles) for the signal in Figure \ref{fig06} in the Hanford and Virgo colored noise. Its standard deviation in each channel is $0.36$ times the maximum amplitude of the signal in the first channel. $\delta_1=\delta_2=0.01$. The color of the circles represents the quantity eq. \ref{3}.}
\label{fig07}
\end{figure}

\begin{figure}[htbp]
\centering\epsfig{file=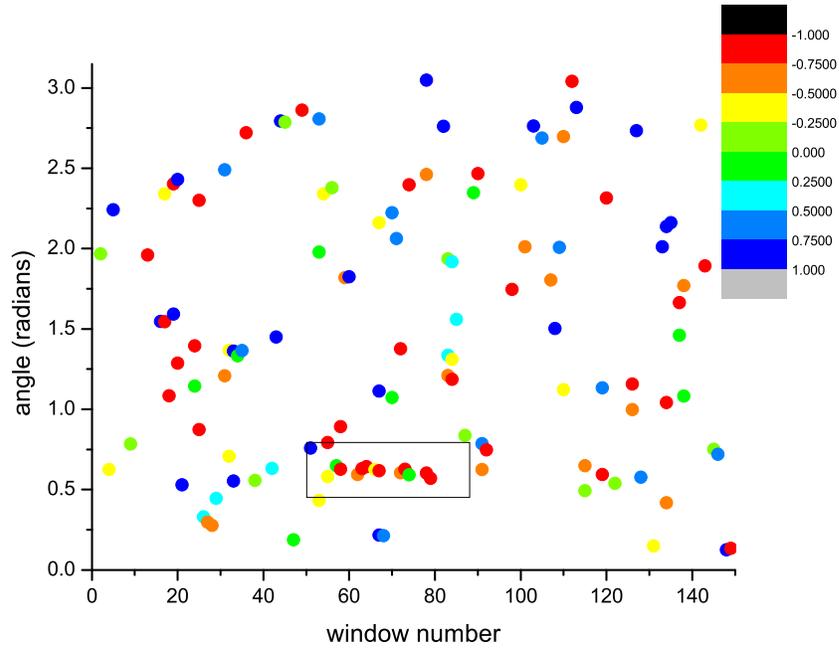,width=0.7\linewidth}
\caption{Coincidences for the signal in Figure \ref{fig01} in white Gaussian noise whose standard deviation in each channel is $1.0$ times the maximum amplitude of the signal in the first channel; $\delta_1=\delta_2=0.04$. The color of the dots represents the quantity eq. \ref{3}.}
\label{fig08}
\end{figure}

Figure \ref{fig08} shows the case of the two signals of Figure \ref{fig01} to which is added white Gaussian noise with standard deviation $1.0$ times the maximum amplitude of the signal: the signal (outlined by a black box) can be recognized only because it's the only sizable sequence of coincidences among many scattered ones. 
   
\begin{figure}[htbp]
\centering\epsfig{file=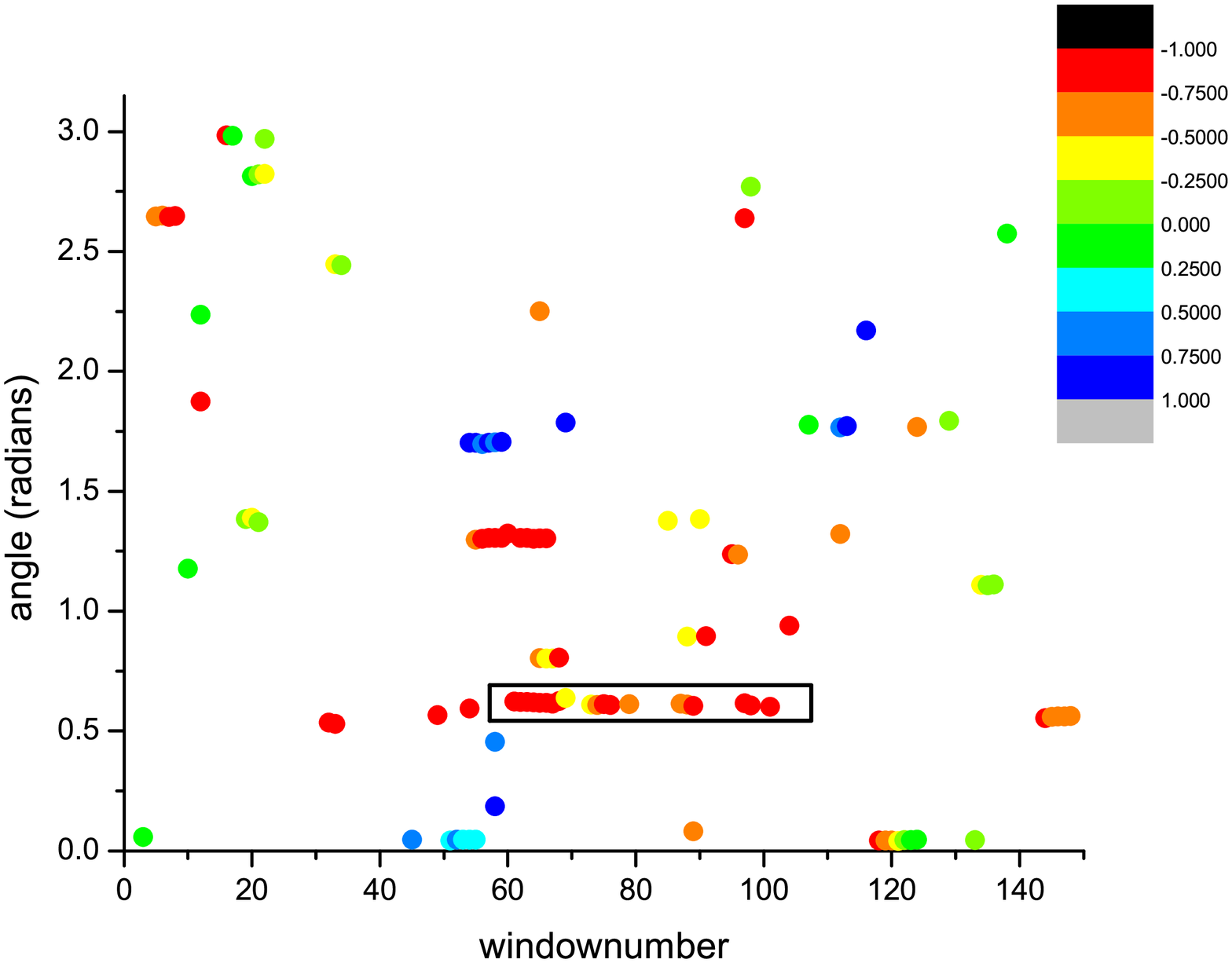,width=0.7\linewidth}
\caption{Coincidences for the signal in Figure \ref{fig01} in the Hanford and Livingston colored noise 
whose standard deviation in each channel is $3.6$ times the maximum amplitude of the signal in the first channel; $\delta_1=\delta_2=0.02$. The color of the dots represents the quantity eq. \ref{3}.}
\label{fig09}
\end{figure}

Figure \ref{fig09} shows the case of the two signals of Figure \ref{fig01} to which is added the Hanford and Livingston colored noise 
with standard deviation in each channel $3.6$ times the maximum amplitude of the signal in the first channel: the signal still gives the longest sequence of coincidences, but some short sequences also appear that are due to the noise. Unless we have an idea of the length in time of the signal we are searching, all these shorter sequences will have to be considered as potential signals. 

\begin{figure}[htbp]
\centering\epsfig{file=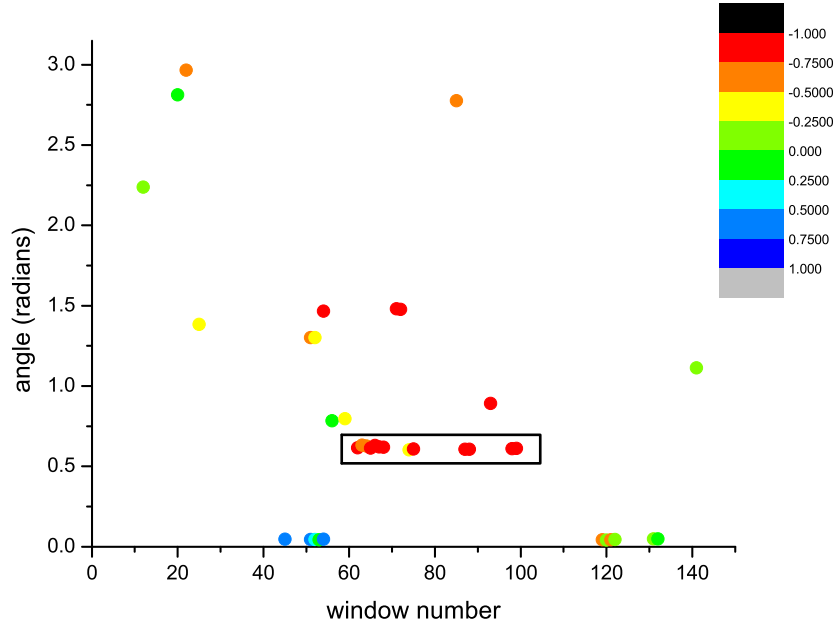,width=0.7\linewidth}
\caption{Same as in Figure \ref{fig09} but we also added white Gaussian noise whose standard deviation in each channel is $0.3$ times the maximum amplitude of the signal in the first channel.}
\label{fig09b}
\end{figure}

\begin{figure}[htbp]
\centering\epsfig{file=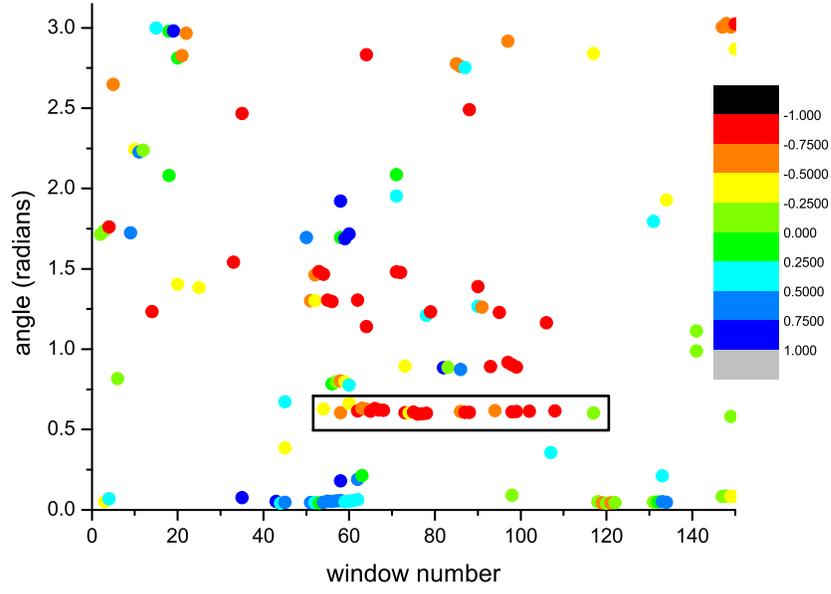,width=0.7\linewidth}
\caption{Same as in Figure \ref{fig09b} but $\delta_1=\delta_2=0.03$.}
\label{fig09c}
\end{figure}

Due to the nonlinearity of the Pad\'{e} transform, adding to the noisy signal a different white Gaussian noise realization at each window step can improve the situation: see Ref. \cite{barone} where different noise realizations are added to the same noisy data set; since we are using a traveling window with short step, the addition of a different white Gaussian noise realization at each window step is enough to break the window to window correlations of the noise poles. Figures \ref{fig09b} and \ref{fig09c} show how the result for Figure \ref{fig09} changes when we add white Gaussian noise with standard deviation $0.3$ times the maximum amplitude of the signal: keeping $\delta_1=\delta_2=0.02$ (Figure \ref{fig09b}) some of the signal points are lost, but all the other sequences have disappeared; if we change $\delta_1=\delta_2$ to $0.03$ (Figure \ref{fig09c}) other scattered coincidences appear but they do not form sequences, while the signal sequence appears enhanced.

\begin{figure}[htbp]
\centering\epsfig{file=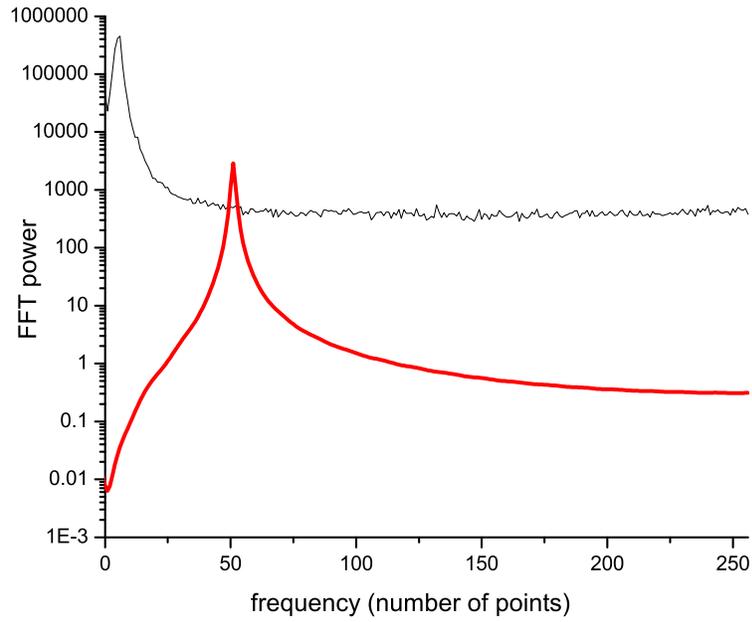,width=0.7\linewidth}
\caption{Noise (black) and signal (red) FFT amplitude spectra; the noise spectrum is averaged over 100 consecutive data segments.}
\label{fig10}
\end{figure}

Note that since the noise is colored and therefore strongly correlated, the Signal to Noise Ratio (SNR) is not the ratio of the total powers $1/3.6$; in this case we can define SNR as
\begin{equation}
SNR=\sqrt{\Sigma_{k=1}^N\frac{|\tilde{h}(f_k)|^2}{<|\tilde{n}(f_k)|^2>}},
\end{equation}
where $\tilde{h}(f_k)$ and $\tilde{n}(f_k)$ are the Fourier transforms of signal and noise respectively, $N$ is approximately the length of the damped signal in number of data points, and $<.>$ represents ensemble average and -assuming the system to be ergodic- is calculated over several time intervals (we use $100$). This SNR gives us an idea of how much of the signal spectrum pokes out of the noise spectrum: see Figure \ref{fig10}. As $N$ is here about $500$ points, the SNR is about $5.5$ for the first channel.  

\begin{figure}[htbp]
\centering\epsfig{file=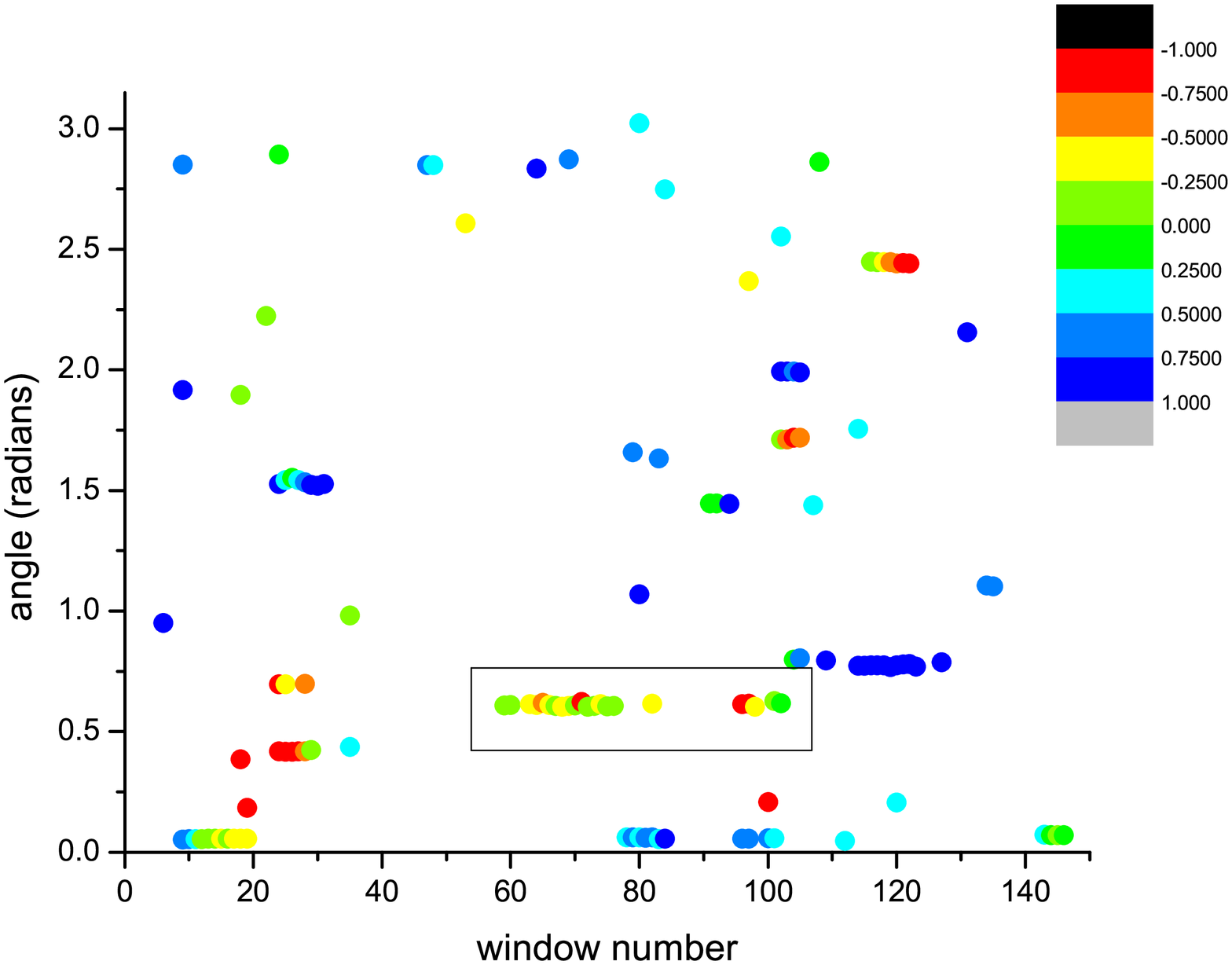,width=0.7\linewidth}
\caption{Coincidences for the signal in Figure \ref{fig06} in the Hanford and Virgo colored noise  
whose standard deviation in each channel is $1.44$ times the maximum amplitude of the signal in the first channel; $\delta_1=\delta_2=0.02$. The color of the dots represents the quantity eq. \ref{3}.}
\label{fig11}
\end{figure}

\begin{figure}[htbp]
\centering\epsfig{file=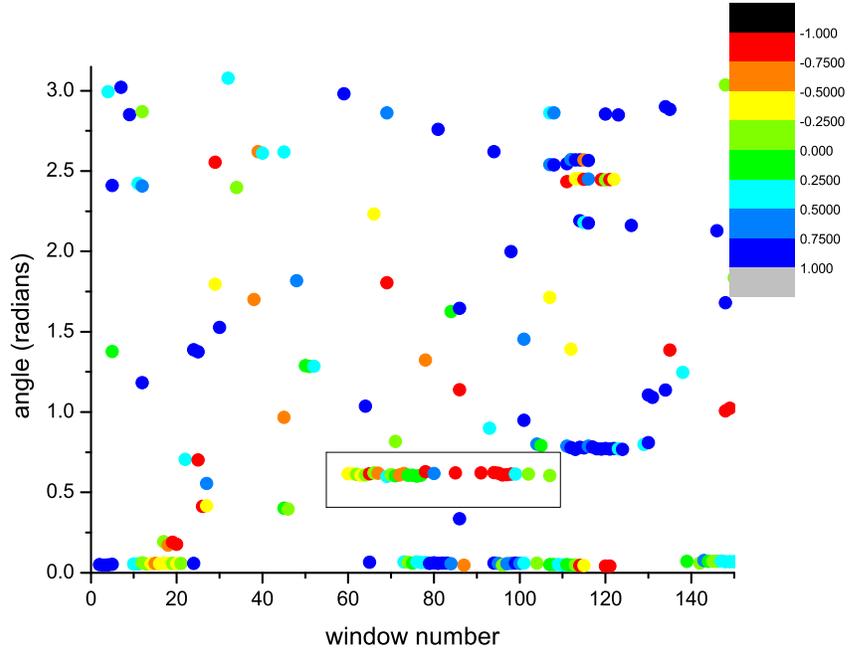,width=0.7\linewidth}
\caption{Same as in Figure \ref{fig11} but we also added white Gaussian noise whose standard deviation in each channel is $0.12$ times the maximum amplitude of the signal in the first channel. $\delta_1=\delta_2=0.03$.}
\label{fig12}
\end{figure}

Figure \ref{fig11} shows the case of the two signals of Figure \ref{fig06} to which is added the Hanford and Virgo colored noise 
with standard deviation in each channel $1.44$ times the maximum amplitude of the signal in the first channel (SNR is about $13.7$ for the first channel and $6.8$ for the second one). As the second signal is significantly smaller than the first one, we had to reduce the noise level to obtain a situation similar to that of Figure \ref{fig09}: again the signal still gives the longest sequence of coincidences, but some significant sequences also appear that are due to the noise. Here too adding some white Gaussian noise appears to somehow improve detection: see Figure \ref{fig12}.  

\section{Conclusions}

We have thus shown the feasibility of detecting the coincident presence of a transient GW signal in the output of two separate antennas through the properties of its Pad\'{e} poles.

We expect our procedure to work at its best for signals having a clearly dominant frequency which is constant or varies slowly as compared to the sampling step. The presence of many frequencies with comparable amplitudes or of sidebands due to the chirping of the main frequency would otherwise drain a significant part of the power from the main pole, thus reducing its residue and confusing the picture.  

To give an idea of the computing times involved, the examples given above ($150$ windows of $100$ points each) take about $3$ minutes each on a desktop computer. 

As a next step, we plan to apply our method to a more realistic noise model including e.g. glitches and to other signal waveforms such as sine-Gaussians or some of the expected core-collapse supernova signals in Ref. \cite{super}. 

At that point a statistical analysis with the evaluation of detection and false alarm probabilities will become possible, which is required to build a full data analysis pipeline.

\section{Acknowledgments}

We thank J.D. Fournier for comments and suggestions.

Special thanks to Professor Mario Diaz, Director of the Center for Gravitational Waves at the University of Texas at Brownsville: without his constant support this work would have never been possible.

Supported by a NASA sub-award to the center for Gravitational Waves, Texas University at Brownsville, Texas USA.

\end{document}